\documentstyle[11pt]{article}
\newcommand{\delg}{{\delta_g}}
\newcommand{\delb}{\bar{\delta}}
\newcommand{\G}{{\cal G}}

\newcommand{\eq}{{equation~}}

\newcommand{\cdm}{{\rm cdm}}
\newcommand{\hdm}{{\rm hdm}}
\newcommand{\bea}{\begin{array}{c}}
\newcommand{\eea}{\end{array}}
\newcommand{\beq}{\begin{equation}}
\newcommand{\eeq}{\end{equation}}
\newcommand{\beqa}{\begin{eqnarray}}
\newcommand{\eeqa}{\end{eqnarray}}

\newcommand{\xibar}{\overline{\xi}}

\newcommand{\gtilde}
	{\mathrel{\raisebox{-1ex}{$\stackrel{\textstyle >}{\sim}$}}}
\newcommand{\ltilde}
	{\mathrel{\raisebox{-1ex}{$\stackrel{\textstyle <}{\sim}$}}}

\newcommand{\lexp}{\mathop{\bigl\langle}}
\newcommand{\rexp}{\mathop{\bigr\rangle}}

\def\etal{{\it et al.\ }}
\def\Or{{\cal O}}

\font\twelveBF=cmmib10 scaled 1200
\newcommand{\x}{\hbox{\twelveBF x}}

\newcommand{\r}{\hbox{\twelveBF r}}
\newcommand{\k}{\hbox{\twelveBF k}}

\def\lim{{\rm lim}}

\def\Mpc{{\,h^{-1}\,{\rm Mpc}}}

\begin{document}
\baselineskip 18pt
\begin{titlepage}
\begin{flushright}
FERMILAB-Pub-93/140-A \\
June 1993 \\
\end{flushright}
\vskip 1.5cm
\begin{center}
{\Large\bf The Three-point Function as a Probe \\ of
           Models for Large-scale Structure}
\vskip .5cm
\end{center}
\begin{center}

{\large\sc Joshua A. Frieman$^{1}$ \&  Enrique Gazta\~naga$^{1,2}$}
\vskip 0.4cm

{\sl $^1$NASA/Fermilab Astrophysics Center \\
Fermi National Accelerator Laboratory \\
Batavia, IL 60510-0500, USA} \\

\vskip 0.3cm

{\sl $^2$Department of Astrophysics \\
University of Oxford\\
Oxford, England OX1 3RH} \\

\vskip .9cm

{\small\rm ABSTRACT}
\end{center}

We analyze the consequences of models of structure formation
for higher-order ($n$-point) galaxy correlation functions in the
mildly non-linear regime. Several variations
of the standard $\Omega=1$ cold dark matter model with scale-invariant
primordial perturbations have recently been introduced to
obtain more power on large scales, $R_p \sim 20$ h$^{-1}$ Mpc, e.g.,
low-matter-density (non-zero cosmological constant) models, `tilted'
primordial spectra, and scenarios with a mixture of cold and
hot dark matter. They also include models with an effective
scale-dependent bias, such as the
cooperative galaxy formation scenario of Bower, etal. (1993).
We show that higher-order ($n$-point) galaxy correlation functions can provide
a useful test of such models and
can discriminate between models with true large-scale power in the
density field and those where the galaxy
power arises from scale-dependent bias: a bias with rapid
scale-dependence leads
to a dramatic decrease of the hierarchical amplitudes
$Q_J$  at large scales, $r \gtilde R_p$.
Current observational constraints
on the three-point amplitudes $Q_3$ and $S_3$ can place limits on
the bias parameter(s) and
appear to disfavor, but not
yet rule out, the hypothesis that scale-dependent
bias is responsible for the extra power observed on large scales.
\bigskip

\bigskip

\noindent{\it Subject Headings}:
Large-scale structure of the universe --- galaxies: clustering

\end{titlepage}

\newpage
\baselineskip 18pt

\section{Introduction}

Recent observations of galaxy clustering in both photometric and
spectroscopic surveys have found more
relative power on large scales, $R_p \sim 20 \Mpc$ ($h=H_0/100$ km/sec/Mpc),
than that expected in the standard cold dark matter (CDM) model of structure
formation (e.g., Maddox, etal. 1990, Efstathiou, etal. 1990,
Baumgart and Fry 1991, Gramann and Einasto 1991,
Hamilton, etal. 1991, Peacock and Nicholson 1991, Saunders, etal. 1991,
Loveday, etal. 1992, Fisher, etal. 1992,
Park, etal. 1992, Vogeley, etal. 1992, Feldman, etal. 1993).
More precisely, the {\it shape} of the observed
galaxy power spectrum $P_g(k)$ or
of its Fourier transform,
the two-point galaxy correlation function $\xi_g(r)$,
differs on these scales from the standard CDM model prediction.

Recall that in the standard CDM model, the Universe is spatially
flat, with a density $\Omega_{cdm} = 1 - \Omega_B \simeq 0.95$
in non-baryonic, weakly interacting particles which have
negligible free-streaming length, and the Hubble parameter $h=0.5$.
Additionally, one posits
that the density perturbations responsible for large-scale structure
are adiabatic and Gaussian, with a scale-invariant primordial power spectrum
$P(k) =\langle|\delta_k(t_i)|^2\rangle \sim k$, as expected in canonical
inflation scenarios. The present spectrum is related to the primordial one
through the transfer function, $T(k;\Omega_i,h)$, which encodes
the scale-dependence of the linear growth of perturbations,
$\langle|\delta_k(t_0)|^2\rangle= T^2(k)\langle|\delta_k(t_i)|^2\rangle$.
Finally, the galaxy power spectrum is related to the density spectrum by a
bias factor $b_g$,
\beq
P_{g}(k)=b^2_g T^2(k)|\delta_k(t_i)|^2 ~~~.
\label{pbt}
\eeq

A number of alternatives have been suggested to remedy the shape
of the CDM galaxy spectrum, each of which involve modifications of
one or more of the standard ingredients of the CDM model in \eq(\ref{pbt}).
These include models with a lower density of cold dark matter,
$\Omega_{\cdm}h \simeq 0.2$, plus a cosmological constant to retain
spatial flatness (Efstathiou, Sutherland, and Maddox 1990),
and models with a mixture of cold and hot dark matter,
$\Omega_{\cdm} \simeq 0.7$, $\Omega_{\hdm} \simeq 0.3$ (e.g.,
Schaefer, etal. 1989,  van Dalen and Schaefer 1992,
Taylor and Rowan-Robinson 1992, Davis, etal. 1992, Pogosyan and
Starobinsky 1992, Klypin, etal. 1992). In these two cases,
the transfer function $T(k)$ is flattened on scales $k^{-1}\sim
R_p$ compared to standard CDM. CDM models with `tilted'
non-scale-invariant, power-law primordial spectra,
$\langle|\delta_k(t_i)|^2\rangle \sim k^n$ with
$n < 1$, which arise naturally in several models of inflation, have
also been recently explored (Adams, etal. 1993, Cen, etal. 1992,
Gelb, etal. 1993, Liddle and Lyth 1992, Liddle, etal.
1992, Vittorio, etal. 1988).
In addition, there is a growing literature on models with non-Gaussian
initial fluctuations; in some cases, initial skewness and/or kurtosis
can lead to enhanced
structure on large scales (e.g., Moscardini, etal. 1993 and references
therein). While such models can display interesting behavior of the
higher order moments, in this paper we will focus on initially Gaussian
fluctuations.

In all these variations on the CDM theme, one important assumption is left
unchanged: that the observable galaxy distribution is related through a simple
bias mechanism to the underlying matter distribution predicted by theory
(e.g., Bardeen, etal. 1986).  In essence, following Kaiser (1984a,b) and
Bardeen (1984), one assumes that galaxies form from peaks above
some global threshold in the smoothed linear
density field. In the limit of high threshold and small variance, this
model is well approximated by
the commonly employed linear bias scheme, in which the galaxy and mass
density fields, $\delta_g(\x)= (n_g(\x)-\bar{n_g})/\bar{n_g}$ and
$\delta(\x)=(\rho(\x)-\bar{\rho})/\bar{\rho}$, are
linearly related through a constant bias factor,
\beq
\delta_g(\x) = b_g \delta(\x)  ~~~.
\label{bg}
\eeq
This relation, implicitly assumed in \eq(\ref{pbt}),
embodies the standard model for biased galaxy formation.

Early numerical evidence for biasing came from the CDM simulations of
White, etal. (1987), which showed that dark matter halos are more
strongly clustered than, and thus `naturally' biased with respect to,
the mass. However, since galaxy formation is a complex, non-linear process
involving both gravitational and non-gravitational interactions,
the relation between the mass and the galaxy distributions may
be more complicated than in the peak bias model. Even purely
gravitational high-resolution N-body simulations suggest that
virialized halos are not always well identified with peaks in the linear
density field (Katz, Quinn, and Gelb 1992).

It is therefore of interest to ask whether a more or less well-motivated
modification of the standard bias scheme
can generate the excess large-scale power within
the context of the standard CDM model. This idea has been
recently studied by Babul and White (1991) and by Bower, etal. (1993) (for
precursors, see Rees 1985, Silk 1985 and Dekel and Rees 1987).
The common thread in these ideas is that the bias mechanism can be modulated
by environment-dependent effects. For example,
in their cooperative galaxy formation scenario, Bower, etal. (1993) (hereafter
BCFW) suggest that the threshold above which perturbations actually
form bright
galaxies may be lower in large-scale, high-density regions than elsewhere.
Or perhaps baryons may be inhibited from cooling in regions photoionized by
an early generation of quasars (Babul and White 1991).
The net result of these feedback mechanisms is that the transformation
from the
density field $\delta(\x)$ to the galaxy field $\delta_g(\x)$ becomes
{\it non-local} (by contrast with \eq(\ref{bg})), and the effective
bias factor becomes scale-dependent. If the bias factor increases
with scale, the galaxy spectrum will have more power at large
scales, as desired. This modification of the standard CDM scenario is
fundamentally different from those mentioned above: with scale-dependent bias,
the extra large-scale power relative to standard CDM
is only apparent, in the sense that it is
only a property of the galaxy field, not the underlying mass density
field; by contrast, in the other CDM variants
(non-zero $\Lambda$, tilt, or mixed dark matter),
there is genuine extra power in the density field.

In this paper, we consider how the higher order irreducible
moments of the galaxy distribution can be used as a test of
models for large-scale structure.
We consider the standard CDM model and its variants with extra large-scale
power (in particular, $\Omega h =0.2$ CDM), as
well as a generalized version of the non-local, scale-dependent bias
scheme embodied in the cooperative galaxy formation (hereafter, CGF)
model of BCFW
in the context of otherwise-standard CDM. Using the results of second-order
perturbation theory (Fry 1984), we compare in detail the
predictions of these models for the three-point function $\xi_3$
with data from the Center for Astrophysics (CfA, Huchra, etal. 1983),
Southern Sky (SSRS, Da Costa, etal. 1991), and Perseus-Pisces
(Haynes and Giovanelli 1988) redshift surveys in
the mildly non-linear regime ($\xi_2 <1$).
Since $\xi_3$ is of second-order in the
density perturbation amplitude for initially Gaussian fluctuations,
for self-consistency we must
generalize the models to include the possibility of non-linear (as well
as non-local) bias and extend them from Gaussian to hierarchical
matter fields.
[We will use the well known result that, at least in the mildly non-linear
regime, the matter field evolved gravitationally from Gaussian initial
conditions leads to hierarchical statistics of the form
$\lexp\delta^J\rexp \propto \lexp\delta^2\rexp^{J-1}$
(cf. Fry 1984, Goroff \etal 1986, Bernardeau 1992).]
The allowance for non-linear bias introduces an additional dimensionless
parameter into the model. Even with this additional degree of
freedom, we find that the CGF model tends to require rather large
values of the bias parameter in order to match the 3-point function
data, because scale-dependent bias modifies the correlation hierarchy,
leading to a dramatic decrease of the hierarchical amplitudes
$Q_J$ at large scales, $r \gtilde R_p$. In the context of standard CDM,
such a high bias is in conflict with the COBE DMR observations
of microwave anisotropy on large scales.  We show that
observations of the 3-point function in Fourier space, $Q(k)$,
on the largest scales accessible to current redshift surveys should
provide a definitive test of the CGF model and
of more general models with scale-dependent bias. Our basic conclusion is
that the scale-dependent bias solution to the problem of extra large-scale
power affects the 3-point functions very differently from
models with genuine extra power (such as CDM with $\Omega h = 0.2$).
Thus, the higher-order correlations provide an important test to
distinguish between different solutions of the extra power problem.

The paper is organized as follows. In section II, since it may
be less familiar to the reader, we briefly
review and generalize the CGF model and recapitulate the results of BCFW,
demonstrating the enhancement of the two-point function on large scales
required to fit the APM angular correlation function data. In
section III, we review the results on the 3-point and higher order
correlations in perturbation theory, focusing on the evolution
of an initial Gaussian density field into a hierarchical field.
In section IV, we study the higher order moments in the CGF
model. Self-consistency demands that we further extend the
model to include non-linear bias. In section V, we compare the
standard CDM, low-density CDM, and CGF-modified CDM predictions
to the data on the 3-point function from the CfA, SSRS,
and Perseus-Pisces redshift surveys and we conclude in section VI.

\section{Cooperative Galaxy Formation and Scale-Dependent Bias}

The cooperative galaxy formation (CGF) model of
BCFW is a simple phenomenological prescription for
obtaining a scale-dependent bias. It
starts with the standard assumptions of the CDM model, but the biasing
mechanism is modified from the high peak threshold scenario.
In the standard peak bias model (Kaiser 1984a, Bardeen, etal. 1986), the
sites of galaxy formation are identified with
peaks of the smoothed linear density field.
That is, one convolves the initial density field with a
filter of characteristic scale $R_g \sim 1 \Mpc$, and
then identifies galaxies with peaks of the smoothed field above
some threshold $\nu \sigma$, i.e., with density maxima satisfying
$\delta(\x_{pk}) > \nu \sigma$,
where $\sigma^2_{R_g} = \langle (\rho - \bar{\rho})^2
\rangle/\bar{\rho}^2$ is the variance of the smoothed field, and $\nu$
sets the threshold height. (Hereafter, we implicitly assume
the field $\delta(\x)$ is smoothed on the scale $R_g$.)
For example, for an infinitely sharp threshold, the galaxy field is
$\delta_g(\x_{pk}) = \theta(\delta(\x_{pk})-\nu \sigma)$.
The combination of
the threshold peak height $\nu$ and the spatial smoothing scale $R_g$ is
chosen so that the density of peaks
reproduces the observed abundance of luminous galaxies;
moreover, these parameters are taken to be global, spatially invariant
quantities.
In the limit of high threshold ($\nu \gg 1$) and small variance,
the two-point correlation function of the peaks is enhanced over that of
the mass by an approximately constant factor (Kaiser 1984a),
\beq
\xi_{pk}(r;\nu) \simeq \left({\nu^2\over \sigma^2}\right) \xi(r) ~~~,
\label{pknu}
\eeq
where $\xi(r) = \langle \delta(\x) \delta(\x+\r) \rangle$.
Since $\xi(r)$ is quadratic in the density field, this is equivalent
to the linear bias model of \eq(\ref{bg}), with the identification
of the bias factor as $b_g = (\nu/\sigma)$. Following Kaiser (1984a)
and BCFW, we will apply this model to regions
above the threshold, $\delta(\x) > \nu \sigma$, rather than to
maxima; this simplifies the model while retaining its important features.

BCFW extend the standard bias model by replacing the universal
threshold $\nu$ with a threshold that depends on the mean mass
density in a surrounding `domain of influence' of characteristic
size $R_s > R_g$. The motivation is to model the possibility that
peaks form galaxies more easily (or perhaps form brighter galaxies
which are included in a magnitude-limited catalog) if there are other peaks
nearby--thus the name cooperative galaxy formation.
Specifically, they assume that galaxies form from regions satisfying
\beq
\delta(\x) > \nu \sigma - \kappa \bar{\delta}(\x;R_s) ~~,
\label{mod}
\eeq
where $\bar{\delta}(\x;R_s)$ is the density field smoothed on the scale $R_s$,
and $\kappa$ is the modulation coefficient of the threshold. If
$\kappa > 0$, the threshold for galaxy formation is lower in
``protosupercluster" regions than in ``protovoids". The parameters
$R_s$ and $\kappa$ parametrize the scale and strength of cooperative
effects; they are also constrained by the observed galaxy abundance.

The model of \eq(\ref{mod})
is equivalent to applying the standard threshold bias model to the new density
field defined by
\beq
\delta'(\x) \equiv\delta(\x)+\kappa\delb(\x;R_s) ~~~,
\label{prime}
\eeq
that is, to imposing the condition $\delta' > \nu \sigma$. Note that
$\delta'$ is a Gaussian random field if the underlying density
field $\delta$ is Gaussian.
Here we consider a generalization of the CGF model: instead of
applying a sharp threshold clipping to $\delta'(\x)$, we assume that
the galaxy field is an arbitrary continuous function of the field $\delta'$,
\beq
\delg(\x)=f(\delta'(\x)) = f\left[\delta(\x)+\kappa\delb(\x;R_s)\right] ~~.
\label{eq:bias}
\eeq
For example, in the limit of high threshold, for the standard
bias model the function $f$ is approximately an exponential, $f(x) =
\exp(\nu x/\sigma)$
(Kaiser 1984b, Politzer and Wise 1984).
We assume that $f$ is expandable in a Taylor series in its argument,
\beq
\delta_g = f(\delta') =\sum_{k=1}^\infty {b_k \over {k!}} {\delta'}^k ~~~.
\label{eq:taylor}
\eeq
BCFW compute the two-point correlation function for the CGF
model on large scales, using the CDM density spectrum
derived from linear perturbation theory.
In this regime, our generalized CGF model reduces to the linear
bias model applied to the field $\delta'$,
\beq
\delg(\x)=b_g \delta'(\x) = b_g\left[\delta(\x)+\kappa\delb(\x;R_s)\right]
\label{lincgf}
\eeq
where we have identified $b_g = b_1$.
That is, by working only to linear order in perturbation theory,
one should self-consistently include only
the first (linear) term in the series of \eq(\ref{eq:taylor}).
Conversely, when we consider second order perturbations below,
we can and should include the possibility of quadratic ($k=2$) bias.

Comparing \eq(\ref{lincgf}) with \eq(\ref{bg}), it is clear that
cooperative effects boost the galaxy power spectrum on large
scales relative to the standard global bias model.
Taking a Gaussian filter for the smoothed density field,
\beq
\bar{\delta}(\x;R_s) = \left(2\pi R^2_s \right)^{-3/2}\int d^3r \delta(\r)
{\rm exp}\left(-{|\x-\r|^2\over 2 R^2_s}\right) ~~,
\label{dsm}
\eeq
the Fourier transforms of the density fields satisfy
\beq
\delta'(\k)=\delta(\k)~\left[1+\kappa \G(\k)\right] ~~~,
\label{ftdk}
\eeq
where $\G$ is the Fourier transform of the window filter
in $\delb(\x;R_s)$,
\beq
\G(\k)= \G(k)= e^{-(kR_s)^2/2} ~~,
\label{Gkdef}
\eeq
with $k=|\k|$.
The galaxy power spectrum, $P_g(k) = \langle |\delta_g(\k)|^2
\rangle$, is thus related to the density power spectrum,
$P(k)= \langle |\delta(\k)|^2\rangle$, by
\beq
P_g(k) = b^2_g P'(k)= b^2_g  \left[1 +\kappa\G(k)\right]^2 P(k) \equiv
b^2_{\rm eff}(k) ~P(k) ~~~.
\label{eq:P'}
\eeq
This expression makes manifest how cooperative effects result in
an effective  scale-dependent bias, $b_{\rm eff}(k) = b_g[1+\kappa\G(k)]$.
On small lengthscales, $k^{-1} \ll R_s$,
\eq(\ref{eq:P'}) implies the usual bias factor, $b_{\rm eff}(k \rightarrow
\infty) \simeq b_g$,
while on large scales, $k^{-1} \gg R_s$, the effective bias
factor is increased to $b_{\rm eff}(k \rightarrow 0) \simeq b_g(1+\kappa)$.
In the parameter range studied by BCFW, the choice
$\kappa = 2.29$, $R_s = 20 \Mpc$ appears to give
the best fit to the observed extra large-scale power for CDM when
compared to the APM angular correlation function, and we shall
focus mainly on this case.
We see that this choice boosts the galaxy power spectrum on
scales $k \ltilde 0.05 h$ Mpc$^{-1}$ by over a factor of ten.

To see what these effects look like graphically for the CDM model, we
consider the linear CDM density power spectrum of Davis, etal. (1985),
\beq
P(k) = A \sigma^2_8 k \left(1+{1.7k\over \Omega h}+{9k^{3/2}\over
(\Omega h)^{3/2}}+{k^2\over (\Omega h)^2}\right)^{-2} ~~~,
\label{cdmpk}
\eeq
where the wavenumber $k$ is in units of $h$ Mpc$^{-1}$. Here
the normalization is set as usual in terms of the variance of the linear mass
fluctuation within spheres of radius 8$\Mpc$,
$\sigma_8 \equiv \langle (\delta M/M)^2 \rangle^{1/2}_{R= 8 h^{-1}
{\rm Mpc}}$, where
\beq
\sigma_R^2 = {1\over 2\pi^2}{\intop_0^\infty} dk k^2 P(k)W^2(kR) ~~,
\label{sigr}
\eeq
and the top-hat window function
\beq
W(kR) = {3\over{(kR)^3}}(\sin kR - kR \cos kR)
\label{wkr}
\eeq
filters out the contribution from small scales.
For standard CDM with $\Omega h = 0.5$,
this gives $A=2.76\times 10^5(\Mpc)^3$.

Substituting the CDM power spectrum with $\Omega h = 0.5$
into \eq(\ref{eq:P'}), we find
the galaxy two-point correlation function for the CGF model
\beq
\xi_g(r) = {1\over 2\pi^2}\int dk~ k^2 {{\rm sin}kr\over kr}P_g(k) ~~,
\label{xicgf}
\eeq
shown in Fig. 1 (the curve labelled CGF, with
$\kappa =2.29$, $R_s = 20 \Mpc$).
Note that we actually plot $\xi_g(r)/(b_g \sigma_8)^2$, where $b_g$
is the constant factor in \eq(\ref{lincgf}). Redshift surveys of optically
selected galaxies (in particular the CfA and Stromlo-APM surveys)
indicate that the variance in galaxy counts on 8 $\Mpc$
scale is of order unity. Thus, in a linear, scale-independent bias model,
the bias factor for these galaxies would be expected to be
$b_{opt} \simeq 1/\sigma_8$; for other
galaxy populations, however, $b_{gal} \sigma_8$ may differ from unity.
For comparison, in Fig. 1 we also show the two-point function for
standard CDM ($\Omega h =0.5$, $\kappa =0$) and for
a low-matter-density CDM model ($\Omega h = 0.2$). Both the CGF
model and the low-density CDM model
have sufficient relative large-scale power to approximately reproduce the
observed galaxy angular correlation function $w(\theta)$ inferred
from the APM survey (BCFW, Maddox, etal. 1990,
Efstathiou, Sutherland, and Maddox 1990). This level of
extra power is also broadly consistent with that inferred from the power
spectrum of IRAS galaxies
(Feldman, etal. 1993, Fisher, etal. 1992) and the redshift-space two-point
function $\xi(s)$ inferred from the Stromlo-APM survey (Loveday, etal. 1992).
The CGF curve in Fig. 1 should be compared to that in Fig. 2 of BCFW.
Note that the linear bias approximation used here (\eq(\ref{lincgf})) differs
from the non-linear threshold formula of BCFW (Cf. their
eqn.(10)), but that our final result for $\xi(r)$ is very similar to theirs.

Thus, cooperative effects can mimic extra large-scale power in the
galaxy two-point function, while the other remedies for CDM, such as
low-density, mixed dark matter, or
tilted ($n < 1$) models, have genuine extra large-scale power in
the spectrum. How can we discriminate between these
choices for extra large-scale power, that is, between real
power and the illusion of power? Below, we argue that the
three-point function can provide a distinguishing test, at least for
models with Gaussian initial fluctuations. The reason is that the
galaxy three-point function induced by gravitational evolution
depends in large measure on the two-point function of the {\it mass}.

Before turning to higher order correlations, we remark that the
treatment given here and below applies more generally than to
the CGF model, and in fact to any model with scale-dependent bias.
That is, the chain of reasoning above is invertible: if
the galaxy and density power spectra are related by a
scale-dependent bias, $P_g(k) = b^2(k) P(k)$, we can always think
of the galaxy field $\delta_g(\x)$ as arising from some
non-local transformation of the density field $\delta(\x)$. To see this, let
$b^2(k)=b^2_g f^2(k)$, where $b_g$ is a constant and we assume that
$f^2(k)$ has a limit, $f^2(k\rightarrow \infty)=1$. Then consider
the field $\delta'(\k)=\delta_g(\k)/b_g=f(k)\delta(\k)$. We can
write $f(k)=1+ \kappa G(k)$, where lim $G(k \rightarrow \infty) = 0$, and
we can choose $\kappa$ such that lim $G(k \rightarrow 0) = 1$,
so that $\delta'(\k) = \delta(\k)[1+ \kappa G(k)]$. Comparing with
\eq(\ref{ftdk}), we see that this expression,
where $G(k)$ is interpreted as the Fourier transform of some window function
(which in general will not be a Gaussian), is all that we need
for the results discussed here and below to go through.
Provided the function $b(k)$ is not too pathological, this
Fourier transform should exist.

\section{3-point correlation function in perturbation theory}

We want to consider how scale-dependent bias, as embodied for example
in the CGF model, affects the higher order correlation functions.
The motivation for this study is that the galaxy three-point function is
observed to scale in a particular way with the two-point function, and
both perturbation theory and N-body simulations show that
this scaling can arise via non-linear gravitational evolution from
Gaussian initial fluctuations. Since scale-dependent bias introduces
a different scale behavior into the problem, we would expect it
to be manifest as a change in the scaling behavior of the higher
order correlations. We will work in the context of second-order
perturbation theory (Fry 1984), the results of which we review here before
discussing how they are modified by scale-dependent bias. The
perturbative approach should be valid in the mildly
non-linear regime, $\delta \ltilde 1$. In the range where they
overlap, the second-order perturbation theory results below for $S_3$ in
standard CDM appear to be quite consistent with
the N-body simulations of Bouchet and Hernquist (1992).

Defining the Fourier transform of the density field,
\beq
\delta(\k) = {1\over V}\int d^3 x \delta(\x) e^{i {\bf k} \cdot {\bf x}} ~~,
\eeq
we consider the two- and three-point functions in $k$-space,
$\lexp \delta(\k_1)\delta(\k_2) \rexp$ and
$\lexp \delta(\k_1)\delta(\k_2)
\delta(\k_3) \rexp$, which are
the Fourier transforms of the spatial two- and three-point
correlation functions $\xi_2(\x_1,\x_2)$ and $\xi_3(\x_1,\x_2,\x_3)$. By
homogeneity and isotropy, the $\k$-space moments are non-zero
only for $\sum \k_i= 0$,
\beq
\lexp \delta(\k_1)\delta(\k_2) \rexp = \delta_{{\bf k}_1+{\bf k}_2,0}P(k_1)~~,
{}~~~\lexp \delta(\k_1)\delta(\k_2) \delta(\k_3) \rexp = \delta_{{\bf k}_1+
{\bf k}_2+{\bf k}_3,0}B(k_1,k_2,k_3) ~~.
\eeq
This defines the power spectrum $P(k)=\lexp |\delta(k)|^2 \rexp$
and the bispectrum $B_{123}=B(k_1,k_2,k_3)$.

Early observations of clustering on small scales (Groth and Peebles 1977)
suggested that the galaxy two- and three-point functions obey a
scaling hierarchy,
\beq
\xi_3(\x_1,\x_2,\x_3)= Q~\left[ \xi_2(\x_1,\x_2)\xi_2(\x_2,\x_3)+
{}~(1 \leftrightarrow 2) + ~(2 \leftrightarrow 3) \right]
\label{def:Q}
\eeq
with $Q =$ constant $\sim 1$, roughly independent of the size and
shape of the triangle formed by the points $\x_1,\x_2,\x_3$.
If the scaling of \eq(\ref{def:Q}) holds exactly, then
the hierarchical 3-point amplitude $Q$ is also related to the
bispectrum by the $\k$-space version of \eq(\ref{def:Q})
(Fry and Seldner 1982),
\beq
Q \equiv { B_{123} \over{ P_1 P_2 + P_1 P_3 + P_2 P_3}} ~~,
\label{def:qk}
\eeq
with $P_i \equiv P(k_i)$. We will consider \eq(\ref{def:qk}) as the
definition of the amplitude $Q$, even if it is not constant.
In the strongly non-linear regime $\delta \gg 1$,
N-body simulations of CDM and power-law spectrum models do seem
to display the approximate shape- and size-indepedence of \eq(\ref{def:Q})
(Fry, Melott, and Shandarin 1993). However, in second-order perturbation
theory in the mildly non-linear regime,
while $Q$ as defined in \eq(\ref{def:qk}) obeys the
scaling with size, it does
depend on the shape of the configuration in $\k$-space.

To calculate the three-point function in the weakly non-linear regime,
one expands the perturbation
equations in powers of $\delta$, $\delta(\x,t) = \delta^{(1)}(\x,t)
+ \delta^{(2)}(\x,t)+...$, where $\delta^{(1)}$ is the linear
solution, and $\delta^{(2)} = {\cal{O}}(\delta^{(1)})^2$ is the
second-order solution, obtained by using the linear solution in the
source terms. For Gaussian initial fluctuations, the three-point
function vanishes to linear order, $\lexp \delta^{(1)}(\x_1) \delta^{(1)}(\x_2)
\delta^{(1)}(\x_3)\rexp = 0$, and the lowest order contribution
to the bispectrum is $B_{123}=\lexp \delta^{(1)}(\k_1) \delta^{(1)}(\k_2)
\delta^{(2)}(\k_3)\rexp + (1 \leftrightarrow 3) + (2 \leftrightarrow 3)$,
with the result
\beq
B_{123}=\left[ {10\over{7}}~+
\left({{\k_1 \cdot \k_2} \over{k_1 k_2}}\right)
\left({k_1 \over{ k_2}}+{k_2 \over{ k_1}}\right)+
{4\over{7}}\left({{\k_1 \cdot \k_2} \over{k_1 k_2}}\right)^2 \right] P_1 P_2+
{}~(1 \leftrightarrow 3) + ~(2 \leftrightarrow 3)
\label{eq:Qm}
\eeq
(Fry 1984). Strictly speaking, this result holds for initially
Gaussian fluctuations in a matter-dominated
universe with $\Omega=1$, but the work of Juszkiewicz and Bouchet (1991)
shows that the dependence of the three-point function on $\Omega$ is
extremely slight. A particular case of importance is that of
equilateral triangle configurations in $\k$-space,
$k_1=k_2=k_3$, for which $Q(k)\equiv Q_\Delta = 4/7$, independent
of $P(k)$. The independence of this result of the power spectrum
makes it a useful quantity for distinguishing gravitational
from non-gravitational (e.g., bias) effects.
(In general, for other configurations or averages over
configurations, there will be a small dependence on $P(k)$.)
In section IV, we will see how this result is modified by constant
and scale-dependent bias, and compare these predictions
with observations.

Another useful and increasingly popular
characterization of the three-point amplitude, which
does depend on $P(k)$, is the hierarchical averaged amplitude $ S_3$,
\beq
S_3(V) ={{\xibar_3(V)}\over{\xibar_2^2(V)}}= {\lexp \delta^3(\x;V)\rexp
\over \lexp \delta^2(\x;V)\rexp^2} ~~.
\label{S}
\eeq
Here $\xibar_2$ and $\xibar_3$ are
the 2-point and
3-point density correlation functions averaged over a window function
$W(\r)$ of characteristic volume $V$:
\beqa
\xibar_2(V) &=& {1\over{V^2}}\int \int d^3r_1d^3r_2~ \xi_2(|r_1-r_2|)
W(\r_1) W(\r_2) \nonumber \\
\xibar_3(V) &=& {1\over{V^3}} \int \int \int d^3r_1 d^3r_2 d^3r_3~
\xi_3(r_1,r_2,r_3) W(\r_1) W(\r_2) W(\r_3)
\label{xibarv}
\eeqa
In comparing with model predictions, it is useful to think of $S_3$
as the ratio of moments of the density field $\delta(\x;V)$ smoothed over the
volume $V$ (Cf. \eq(\ref{S})),
\beq
\delta(\x;V)={1\over V} \int d^3r~ \delta(\x+\r) W(\r) ~~.
\eeq
Thus, $\xibar_2(V)$ is just the variance of the smoothed density field,
given by \eq(\ref{sigr}), and $\xibar_3(V)$ is its skewness. (The
smoothing discussed here should not be confused with the smoothed
density field introduced in the CGF model of \eq(\ref{prime}); in
the CGF model, the smoothing radius is associated with the physical
scale of threshold modulation effects, while here it merely defines the
resolution with which one observationally probes the density field.)

Following standard practice, we evaluate $S_3$ with a top-hat window:
for the volume $V=4\pi R^3/3$,
$W(r) = 1$ for $r<R$ and vanishes for $r>R$; its Fourier transform $W(kR)$
is given by \eq(\ref{wkr}). In this case, $\xibar_2$ and $\xibar_3$ are
related to the moments of counts in cells of volume $V$, and
the skewness is given by
\beq
\lexp \delta^3(R)\rexp = {3\over (2\pi)^6}\int \int d^3k_1 d^3k_2
B(k_1,k_2,|{\bf k}_1+{\bf k}_2|) W(k_1R) W(k_2R) W(|{\bf k}_1+{\bf k}_2|R) ~.
\eeq
In Fig.2, we plot $S_3$
as a function of the top-hat smoothing radius $R$ for CDM power
spectra with $\Omega h = 0.5$ and $\Omega h = 0.2$
(Cf. (\ref{cdmpk})), using the second-order perturbation theory result
(\ref{eq:Qm}) for the bispectrum (and assuming that the smoothing
radius $R$ is much larger than the galaxy smoothing radius $R_g \sim 1
\Mpc$).
In computing $S_3$ for the low-density model, we have ignored the
tiny correction for $\Omega \neq 1$ (Juszkiewicz and Bouchet 1991).
The result for the CGF model, also
shown here, will be discussed below in section IV.
So far as we are aware, these
numerical results for $S_3$ for CDM are new. (Goroff et al. 1986
roughly integrated $S_3$ for CDM with a Gaussian smoothing window using
Monte Carlo integration, and we have also studied $S_3$ for Gaussian
smoothing. Top hat smoothing requires a more accurate numerical
integrator, and we have checked our integration code by comparing
with the analytic results of Juszkiewicz and Bouchet (1991) for $S_3$ for
power law spectra--see below).
Where our results
overlap with the N-body results of Bouchet and
Hernquist (1992), the agreement is quite good. We see that
$S_3$ does vary with
scale $R$ in a manner that depends on the shape of the power spectrum,
because the CDM spectrum is not exactly scale-free.
For a scale-free, power-law spectrum $P(k) \propto k^n$, $R$ can be scaled out
of the expression for $S_3$, i.e.,
$S_3$ is a constant, and its value can be found analytically,
$S_3(R)=34/7-(n+3)$ (Juszkiewicz and Bouchet 1991). On the other hand, for a
purely unsmoothed field, $R = 0$, $W(kR) = 1$, the normalized skewness is
$S_3(0) = 34/7$, independent of the power spectrum (Peebles 1980).

The hierarchical behavior of the three-point function in perturbation theory
extends to higher order correlations,
so one can define higher order hierachical amplitudes
$Q_J \simeq \xi_J/\xi_2^{J-1}$ or $S_J=\xibar_J/\xibar_2^{J-1}$
which have characteristic amplitudes set by gravitational instability
(see Peebles 1980, Fry 1984, Goroff et al. 1986, Bernardeau 1992).

\section{Scale-dependent bias and the 3-point correlations}

We now turn to study how the $J$-point correlation
amplitudes, and in particular the three-point function, are
affected by constant and scale-dependent biasing.
Because we consider the 3-point function $\xi_3$, we must
extend the CGF model to the case in which the matter distribution is not
just Gaussian but hierarchical, i.e., we consider the contribution of
second-order gravitational evolution. Fry and Gazta\~naga (1993a) have shown
that the first-order contribution of biasing to $\xi_3$ is comparable
to the contribution from second-order gravitational evolution and, thus,
it is not consistent to assume a purely Gaussian density field.
We first consider how the non-local cooperative modulation of the
density field affects the 3-point function, and then study how
it is further affected by linear and non-linear bias, that is,
we consider the sequence of transformations $\delta \rightarrow \delta'
\rightarrow \delta_g$.

\subsection{Cooperative bias}

Consider the effect on the 3-point amplitude of the non-local,
cooperative linear transformation of the density field given in
\eq(\ref{ftdk}).
The bispectrum of the cooperative field $\delta'(\x)$ is
\beq
B'_{123}=B_{123}~(1 +\kappa\G_1)(1 +\kappa\G_2)(1 +\kappa\G_3) ~~~,
\label{eq:B'}
\eeq
where $\G_i \equiv \G(k_i)$ is given by \eq(\ref{Gkdef}).
The hierarchical 3-point amplitude $Q'$ of the field $\delta'$,
defined in
\eq(\ref{def:qk}), can be expressed in terms of the 3-point amplitude $Q$
for the underlying density field, $\delta$:
\beq
Q'
=  Q ~{(P_1 P_2 + P_1 P_3 + P_2 P_3)~(1 +\kappa\G_1)(1 +
\kappa\G_2)(1 +\kappa\G_3) \over{ P_1 P_2 (1 +\kappa\G_1)^2(1 +\kappa\G_2)^2+
{}~(1 \leftrightarrow 3) + ~(2 \leftrightarrow 3)}} ~~.
\label{eq:Q'}
\eeq
Note that the ratio $Q'/Q$ has no explicit angular dependence in $\k$-space,
i.e., it depends only on the magnitudes $k_1$, $k_2$, $k_3$. Using this
property, we can point to several important limiting behaviors of $Q'/Q$.
For example, on small length scales, $k_1$, $k_2$, $k_3 \gg R^{-1}_s$, we
obviously retrieve $Q'=Q$, and in the opposite limit
of large scales (small triangles in $\k$-space),
$k_1$, $k_2$, $k_3 \ll R^{-1}_s$, we have $Q'/Q \simeq 1/(1+\kappa)$,
independent of the power spectrum and the triangle configuration. The other
limiting case of interest is a triangle with two large sides and one small
side, e.g., $k_1$, $k_2 \gg R^{-1}_s$, $k_3 \ll R^{-1}_s$: if the
power spectrum is approximately a power law, $P(k) \propto k^n$, then
for $n>0$ (and $k_3/k_{1(2)} \ll (1+\kappa)^{-2/n}$), $Q'/Q \simeq
1+\kappa$; for $n=0$, $Q'/Q = 3(1+\kappa)/[1+2(1+\kappa)^2]$; and
for $n<0$, $Q'/Q \simeq 1/(1+\kappa)$.

As noted in section III, an important class of configurations is
equilateral triangles in $\k$-space, $k_1 = k_2 = k_3 = k$, for which
\beq
Q_\Delta'= Q_\Delta~ {{(1 +\kappa\G)^3}\over{(1 +\kappa\G)^4}} =
{Q_\Delta\over{(1 +\kappa\G)}} ~~.
\label{Qeq}
\eeq
With the Gaussian CGF smoothing window, $\G(k)= e^{-(kR_s)^2/2}$,
for scales larger than $R_s$, $kR_s \ll 1$,
we have $Q_\Delta'= Q_\Delta~(1+\kappa)^{-1}$, whereas for scales
smaller than $R_s$, $kR_s \gg 1$, we have $Q_\Delta'=Q_\Delta=4/7$.
For the preferred parameter values considered by BCFW
to match the APM data, $R_s=20 \Mpc$ and $\kappa = 2.29$,
we see that within the range of the weakly non-linear regime,
$k^{-1} \sim 10 \Mpc$, there is a sharp transition
from $Q_\Delta' \simeq Q_\Delta$ to $Q_\Delta'= 0.3Q_\Delta$. We explore the
observational consequences of this behavior in the next section (see Fig. 3).

It is also of interest to study the normalized skewness of the
smoothed cooperative density field, $S'_3(R)= \lexp \delta'^3(\x;R)\rexp/\lexp
\delta'^2(\x;R)\rexp^2$, where the Fourier-transform of the
top-hat-smoothed cooperative density field
is $\delta'(\k;R)=W(kR)\delta(\k)[1+\kappa \G(kR_s)]$, with $W(kR)$ given by
(\ref{wkr}) and $\G(kR_s)$ given by (\ref{Gkdef}). The function
$S'_3(R)$ is shown, for
the CDM $\Omega h =0.5$ spectrum with the canonical CGF parameters
$\kappa = 2.29$, $R_s = 20 \Mpc$, as the curve labelled CGF in Fig. 2.
As expected, in this case $S_3$ has a steeper dependence on $R$ for
scales $R \ltilde R_s$ than either the standard or low-density CDM
models, due to the rather sharp, non-power-law
feature in $\delta'(\k)$ arising from cooperative effects. These
different behaviors are compared with data in Fig. 4 below.

One can also define
higher order amplitues, $Q_J$,
by $\lexp\delta^J(k)\rexp \simeq Q_J P^{J-1}$,
  with $Q_3=Q$. From the above arguments it is straightforward to show
that, in general, for regular $J$-sided polygons in $\k$-space,
\beq
Q_J' = { Q_J \over{(1 +\kappa\G)^{J-2}}}
\eeq
Again, for a Gaussian filter $\G$, we have $Q_J'= Q_J~(1+\kappa)^{-J+2}$
for $kR_s \ll 1$ and $Q_J'=Q_J$ for $kR_s \gg 1$.
Thus if the underlying density field $\delta(\x)$ is hierarchical,
with $Q_J$ approximately
constant as a function of $k$, the new field $\delta'$ is also
hierarchical, $\lexp\delta^{'J}\rexp = Q_J' \lexp\delta^{'2}\rexp^{J-1}$, with
the hierarchical amplitudes $Q_J'$ varying with scale from $Q_J'=Q_J$
to $Q_J'= Q_J~(1+\kappa)^{-J+2}$.

\subsection{Non-linear, local bias}

In the previous subsection, we considered the three-point function for
the cooperative density field $\delta'(\x)$ defined in \eq(\ref{prime}).
We now want to relate this to the three-point function of the
galaxy field $\delta_g(\x)$, defined by the arbitrary, local, non-linear
transformation of the cooperative field in \eq(\ref{eq:bias}).
Fry \& Gazta\~naga (1993a) have shown that, in the weakly non-linear limit
$\lexp\delta^2\rexp <1$, the hierarchical relation between the
moments of the density field,
$\lexp\delta^j\rexp\propto\lexp\delta^2\rexp^{j-1}$,
is preserved under an arbitrary local transformation of this form.
Nevertheless, the higher order moments of the galaxy field will differ
quantitatively from the hierarchical amplitudes of the cooperative field.
The analysis of Fry \& Gazta\~naga (1993a) is valid as long as the
amplitudes of the original field (here, the cooperative field $\delta'(\x)$)
are of zeroth order in the two-point function
$\xi_2=\lexp\delta'^2\rexp$, i.e.,
under the assumption that $Q_J' =\Or \lexp\delta'^2\rexp^0$; in particular,
their results apply even if the original field is not hierarchical in
the strict sense that $Q_J'$ is constant.

Let the hierarchical amplitudes of the smoothed galaxy field
$\delg(\x)=f(\delta')$ be denoted by
$Q_{g,J}$. To consider the 3-point galaxy amplitude,
$Q_g \equiv Q_{g,3}$, we must keep terms up to quadratic order in $\delta'$ in
the expansion (\ref{eq:taylor}) of the biasing function $f(\delta')$.
Applying the results of Fry and Gazta\~naga (1993a), we find
\beq
Q_g = b^{-1}( Q' + c_2) + \Or\lexp\delta'^2\rexp ~~~,
\label{Qgtot}
\eeq
where $c_2=b_2/b$ and $b=b_1$ in \eq(\ref{eq:taylor}),
and $Q'$, the 3-point amplitude for the cooperative field $\delta'$,
is related to the 3-point amplitude of the underlying density
field by \eq(\ref{eq:Q'}). For example,
for the high peaks model, in the limit $\nu \gg 1$ and $\sigma \ll 1$,
the bias function $f(\delta')$ is exponential, and
we have $c_2=b$. This suggests that the
$c_2$ term in (\ref{Qgtot}) is of the same order as the $Q'$ term, i.e.,
that the contribution of non-linear bias to the
galaxy 3-point function may be comparable to the second-order gravitational
contribution. For equilateral triangles in $\k$-space,
we can use (\ref{Qeq}) and (\ref{Qgtot}) to relate the galaxy 3-point
amplitude $Q_{g,\Delta}$ to that of the underlying density field,
$Q_\Delta$,
\beq
Q_{g,\Delta} = b^{-1}\left( { Q_\Delta \over{1 +\kappa\G}} +  c_2\right) ~~~,
\label{eq:Qg}
\eeq
with $Q_\Delta=4/7$. On small scales, $k_{NL} \gg k \gg R_s^{-1}$, where
cooperative effects are negligible, but still in the
mildly non-linear regime
$\lexp\delta^2\rexp < 1$, we have $Q_{g,SS} \simeq b^{-1}( Q + c_2)$, just
the result in the absence of cooperative effects. On large scales,
$k \ll R_s^{-1}$, the galaxy 3-point amplitude is
$Q_{g,LS} \simeq b^{-1}[(1+\kappa)^{-1}Q + c_2]$.
The fractional change in $Q_g$ between large ($kR_s<1$) and small ($kR_s>1$)
scales is thus
\beq
{\Delta Q_g \over{Q_g}} \simeq \left({\kappa \over{1+\kappa}}\right)
{Q\over{b Q_g}} ~~~.
\label{coop:q3}
\eeq
For the case of purely linear biasing, $c_2=0$, this gives $\Delta Q_g/Q_{g,SS}
\simeq \kappa/(1+\kappa)$, independent of $b$ or $Q$.

A similar expression can be derived for the hierarchical amplitudes
$S_J=\xibar_J/\xibar_2^{J-1}$ of the volume-averaged correlation functions.
Following the arguments above,
the small-to-large-scale variation in the galaxy 3-point amplitude
$S_g \equiv S_{g,3}$ is related to the density amplitude $S \equiv S_3$ by
\beq
{\Delta S_g \over{S_g}} \simeq \left({\kappa \over{1+\kappa}}\right)
{S\over{b S_g}} ~~.
\label{coop:s3}
\eeq
Again for purely linear bias, this gives
${\Delta S_g/{S_{g,SS}}} \simeq \kappa/(1+\kappa)$.

Finally, for the non-CGF models, note that \eq(\ref{Qgtot}) relates
the galaxy and matter density 3-point amplitudes with the replacement
$Q' \rightarrow Q$ (Fry and Gazta\~naga 1993a); we will make use of this
in comparing the CDM models to observations below. It is also worth
reiterating that all of our results apply to models with initially
Gaussian fluctuations. For non-Gaussian models, there is an additional
first-order contribution to the 3-point amplitude, which can be thought of
as a (possibly scale-dependent) contribution to the parameter $c_2$.

\section{Comparison with observations}

We now compare the model predictions for the three-point amplitudes with
observations from the CfA, SSRS, and Perseus-Pisces redshift surveys.
As above, we focus on three models: standard CDM ($\Omega h =0.5$),
low-density CDM ($\Omega h = 0.2$), and CGF-modified standard
CDM, and we employ the results of second-order perturbation theory.
Comparison with the observed galaxy amplitudes, $S_{g,3}$ and $Q_{g,\Delta}$,
can in principle be used to constrain the bias parameters $b$ and $c_2$
as well as the CGF parameters $\kappa$ and $R_s$. In combination with
other observations, e.g., of the galaxy power spectrum, and of the
large-angle microwave anisotropy as seen by COBE DMR and other experiments,
these results can help point toward preferred models for large-scale
structure.

\subsection{Limits from $Q_{\Delta}$}

Baumgart and Fry (1991)
have estimated the galaxy power spectrum and the Fourier-space three-point
amplitude for equilateral triangle configurations, $Q_\Delta$, using data
from the Center for Astrophysics and Perseus-Pisces redshift surveys.
It is worth noting that the power spectrum
$P(k)$ for these samples does show evidence for the extra large-scale
power inferred in other spectroscopic (e.g., IRAS)
and photometric (e.g., APM) surveys.
Their results for $Q_\Delta(k)$, averaged over 3 subsamples each
from the CfA and Perseus-Pisces surveys, are shown
in Fig. 3. The errors bars in each bin indicate the variance between
subsamples, and
we only show results for values of the wavenumber away from the
strongly non-linear regime.

The striking feature of these results is the relative constancy of
the three-point amplitude over more than a decade in wavenumber,
$k = 0.1 - 1.6 (\Mpc)^{-1}$. Moreover, the observed amplitude of $Q_\Delta$
over this range is apparently in reasonable agreement with the prediction of
second-order perturbation theory {\it without} cooperative effects,
and under the assumption of
no bias, $b = 1$, $c_2 = 0$, namely $Q_\Delta = 4/7$ (shown as the
short-dash line in Fig. 3). Turning this around, using the
perturbation theory
relation $Q_{g,\Delta} = b^{-1}[(4/7) + c_2]$, one can in principle
use the results in
Fig. 3 to constrain the parameter space of $b-c_2$ for any model with
scale-independent bias. In practice, however, the derived constraints
are not terribly stringent. First,
searching this two-dimensional space and treating the data
points as independent, one finds a minimum $\chi^2 \simeq 25.5$
(for 12 data points and a 2-parameter fit, i.e., 10 degrees of freedom)
for $c_2 \simeq 0.52b-(4/7)$. This is consistent with a mean value
of $Q_\Delta \simeq 0.52$ over the plotted range of $k$, close to the expected
perturbation theory result of $4/7=0.57$. In particular, for purely
linear bias, $c_2=0$, the best fit value of the bias parameter is
$b=1.1\pm 0.1$, consistent with the visual impression from Fig. 3.
On the other hand, this constraint on the bias parameter space
should be interpreted with a great deal of caution, since the
best fit curve for perturbation theory has a chi-squared of
2.5 per degree of freedom, more than 3-$\sigma$ above the
expected value. A better fit to the data would be obtained with
a model in which $Q_\Delta$ falls gently with increasing $k$. However,
given the likelihood that the true data errors are larger
than those shown here, it would certainly be premature to exclude the
perturbation theory result on this basis.

The statements above apply for local, non-cooperative bias models.
On the other hand, as noted in section 4.1, the CGF model predicts
a dramatic scale-dependence of $Q_\Delta (k)$ around the scale
$kR_s \sim 1$. This behavior is shown in Fig. 3 for the 3 CGF parameter
choices considered by BCFW, $\kappa, R_s = 0.84, 10 \Mpc$ (dot-long
dash curve), 2.29, $20 \Mpc$ (solid curve), and 4.48, $30 \Mpc$
(dot-short dash curve). As above, these models are plotted for
$c_2=0$, $b=1$. The `smoking gun' of these models is the sharp
downturn in $Q_\Delta$ on large scales. Since, within the
observational errors, no such downturn is observed, one can use
this to constrain the CGF parameter space. In particular, for
$R_s = 10 \Mpc$, $\kappa =0.84$, the CGF model is always a significantly
poorer fit to the data than the scale-independent bias models. For this
choice of CGF parameters, the requirement of a fit that is within
1-$\sigma$ of the scale-independent models (i.e., a fit with
$\chi^2 < 3$ per degree of freedom) necessitates a linear bias parameter
$b > 2.6$ and a significant non-linear bias, $c_2 > 0.8$. In this case,
the large linear bias factor suppresses the gravitational and cooperative
contribution to $Q$, and the match with the observations is obtained
chiefly by the non-linear bias. This would make the apparent
agreement between the observed $Q_\Delta$ and the perturbation theory
prediction of 4/7 purely coincidental. This behavior is an instance
of our general conclusion that models with sharply varying scale-dependent
bias are forced to uncomfortably large values of the linear bias $b$.
On the other hand, for larger values of the CGF `scale of influence'
$R_s$, the 3-point data do not extend to large enough scales for
the downturn to be significant. Consequently, the $R_s = 20 \Mpc$
CGF model, when fitted to the $Q_\Delta (k)$ data, occupies the same
region in the two-dimensional $b-c_2$ bias parameter space, with only
a slightly higher $\chi^2$ than the non-CGF models. Clearly,
to more strongly constrain or rule out the CGF model, it would be useful to
have data on $Q_\Delta (k)$ which extends down to $k \ltilde 0.05$
h Mpc$^{-1}$; this should be feasible with currently available redshift
samples drawn from the IRAS catalog.

\subsection{Limits on $S_3$}

To compare model predictions to observations of the volume-averaged
normalized skewness $S_3$,
we use the results of the $S_3$ analysis by Gazta\~naga (1992) for
samples in the CfA and SSRS redshift catalogs (we use the largest
samples, denoted SSRS115 and CfA92 in
Gata\~naga 1992). The average over these samples is shown in Fig. 4,
where we plot $S_{g,3}$ as a function of top-hat smoothing radius
(or cell size) $R$.
Each data point in Fig. 4 is an average over bins that correspond
to different degrees of freedom: for a given value of $R$,
the average number of galaxies in that cell size
is at least one unit larger than in the cell of the next smallest value of $R$
shown in the figure.
The errorbars shown here are the larger of the intersample dispersion
in the given $R$-bin and the intrinsic errors in the original samples.
 From Fig. 4, it is apparent that $S_{g,3}(R) \simeq 2$ is quite constant over
the range of $R$ shown, with a variation of
about 25\%. We also show in Fig. 4 the same model predictions for $S_3$
as in Fig. 2, again for the bias parameters $b=1$ and $c_2=0$.
The reader should mentally note
that the curves in Fig. 4 can be shifted vertically, and have their
slopes magnified or depressed, by
changing the values of $b$ and $c_2$.

In Fig. 5 we plot the contours of $\chi^2$ for the comparison between
the three models and the $S_3$ observations in the $b-c_2$ parameter space.
The 3 contours correspond to $\chi^2=5$, 8, and 14 for
$11$ data points fit with $2$ parameters (9 degrees of freedom).
Again, because of the way error bars have been assigned to the
data, we caution against absolute interpretations of these $\chi^2$
values; however, the difference in $\chi^2$ values for different models
should provide a measure of the relative goodness of fit to the data.
In this sense, both CDM models give comparably good fits to the data for values
of the bias parameter above $b=1$, with $b>1.8$ for the best fit
(lowest $\chi^2$ contour) range.
For a given value of $b$, the non-linear bias $c_2$ is
slightly larger for the $\Omega h=0.2$ case than for standard
$\Omega h = 0.5$ CDM. For the CGF model, on the other hand,
the linear bias parameter must satisfy $b>2$ for
a reasonable fit, while the best fit range requires $b > 3$.
The large value of bias for the CGF model inferred from $S_3$ is
qualitatively similar to the result above from the Fourier-amplitude
$Q_\Delta$: fits to the data with large values of $b$ are in a sense
{\it ad hoc}, because the agreement is
obtained by depressing the gravitational
contribution and then fitting with the non-linear bias $c_2$ alone.
In particular, for $c_2/b \simeq 0.6$,  the galaxy amplitude
$S_{g,3} \simeq 2 \simeq 3Q$ is completely produced by non-linear bias, not by
gravitational or cooperative effects.
Therefore the fit for the CGF model,
for which $c_2/b=0.6$, does not really reflect agreement between the data and
the CGF model, but rather the possibility that, in any model, the
observed signal comes from the non-linear component of biasing.

At this point, it is worth noting several features of the $S_3$
observations. The skewness has been measured from other redshift and
angular catalogs in addition to those used above;
a useful compendium of results
in the literature is given in Fry and Gazta\~naga (1993b). Except for
the Lick catalog, the values of  $S_3$ inferred from other surveys are
broadly consistent with those shown in Fig. 4
(e.g., Bouchet, etal. 1991, 1993,
Meiksin, etal. 1992). A second issue concerns
redshift distortions of the higher order moments. It is well known
that peculiar velocities distort the galaxy power spectrum (Kaiser 1987),
so that the power measured in a redshift catalog does not precisely
represent the clustering power in real space. The transformation
from the real space to redshift space power spectrum depends on
the ratio $\Omega^{0.6}/b$. The extent to which
this affects higher moments has been somewhat controversial: in N-body
simulations, Lahav
etal. (1993) find that $S_3$ is significantly distorted in redshift
space in the strongly non-linear regime, while Coles, etal. (1993) do
not see this affect. In their analysis of higher moments in the
CfA, SSRS, and IRAS 1.9 Jy catalogs, Fry and Gazta\~naga (1993b)
find that the volume-average 3-point function $\xibar_3$ is affected
by redshift distortions, but that the normalized skewness $S_3$ is
insensitive to them. This empirical insensitivity justifies our
comparison of the model results to the $S_3$ data in redshift space.

We finish this section with some comments about the implications
of the $Q$ and $S_3$ observations for the bias parameter(s) and how
these compare with other data on large-scale structure. We will focus
on the CDM models (as opposed to the CGF model). First, as noted
above, the $Q_\Delta$ observations do not significantly constrain
the bias parameter $b$ once one allows for non-linear bias (although
they do imply a relation between $b$ and $c_2$). On the other hand,
the $S_3$ observations do appear to favor larger values of the
bias, $b \gtilde 1.8$, for both CDM models. In a simple bias
prescription, for CfA galaxies we would expect $b \sigma_8 \sim 1$,
so that, taken at face value, this constraint on $b$ would imply
a low normalization amplitude for the CDM models, $\sigma_8 \ltilde 0.56$.
For standard $\Omega h = 0.5$ CDM, this is uncomfortably low
compared to the amplitude inferred from the COBE DMR measurement of
the large-angle microwave anisotropy, $\sigma_{8,dmr} \sim 1$.
For the low-density CDM model, the $\sigma_8$ amplitude inferred
from COBE has a large range, depending on the choice of $\Omega$
and $h$ (Efstathiou, Bond, and White 1992). For example, for the
choice $\Omega = 0.3$, $h=2/3$, Efstathiou, Bond, and White (1992)
infer $\sigma_8 \sim 0.7$ from COBE, closer to the range implied by the $S_3$
observations. (On the other hand, for lower $\Omega$ and larger $h$,
e.g., $\Omega=0.2$ and $h=1$,
the COBE value for $\sigma_8$ becomes larger than unity, which is
disfavored by the 3-point data.)
While it is tempting to draw conclusions about the viability of
different models from this comparison, in particular, to argue against
standard COBE-normalized CDM, there are potential pitfalls which
mitigate against making high confidence-level statements of this type.
In particular,
if one focused only on the $S_3$ data in Fig.4 at large $R$ (where the
perturbation result is more trustworthy), one would conclude that
standard $\Omega h = 0.5$ CDM fits the data well with
$b \simeq 1$, $c_2 \sim 0$, in agreement with the COBE normalization.
A conclusion which can be drawn with more confidence from Fig. 5 is
that the high peaks model prediction $c_2/b = 1$ is inconsistent with
the $S_3$ data for any of the Gaussian models we have studied.

\section{Conclusion}

We have studied the three-point galaxy correlations in models of
large-scale structure, focusing on the CDM model and its variants
with extra large-scale power, working in the context of biased
galaxy formation in second order perturbation theory.
In the non-local bias scheme, galaxies form and
light up in just such a way as to create the illusion of extra power.
We have shown that models with effective scale-dependent
(or non-local) bias,
such as the CGF model, can display the same enhanced large-scale
power as other variations of standard CDM, but that they break
the scaling hierarchy between the two- and three-point functions
that arises from gravitational evolution. The resulting step
in the Fourier-space three-point function $Q_\Delta (k)$
at the scale $k \sim R^{-1}_s$ of the bend in the bias function
(which produces the extra large-scale power) should
provide a strong observational test of scale-dependent bias models.
However, this step can be partially masked if $b$ is large and if
there is significant non-linear bias. Consequently,
using data currently available, we have shown that the scale-dependent
bias explanation of large-scale power requires a larger value of
the linear bias factor $b$ than in the standard CDM model, and
a substantial non-linear bias, in order to account for the observed
flatness of the three-point amplitudes.

On the other hand, the three-point amplitudes $S_3$ and $Q$ do not
strongly discriminate between standard and low-density CDM with
scale-independent bias; this
conclusion also extends to the tilted CDM and mixed dark matter
models. In these cases, however, the $S_3$ data tentatively
point to moderately large values of the bias, $b \gtilde 1.8$,
but more data on large scales is needed to confirm this.
We emphasize that it is useful to have observational tests
using both $S_3$ and $Q_\Delta$, since the former depends on the
power spectrum while the latter does not.

For completeness, we note that the CGF and other non-local bias
models have other hurdles to overcome in addition to the higher moments.
In the CGF and related models, the effective bias factor increases with
lengthscale. On the other hand, recent N-body simulations of CDM incorporating
hydrodynamics suggest that
the bias factor $b(k)$ {\it decreases} with lengthscale (Cf.
Katz, Hernquist, and Weinberg 1992, Fig.2 of Cen and Ostriker 1992).
In addition, the modifications introduced by CGF do not apparently
address the difficulties which CDM faces with excessive
pairwise velocities on small scales
(Gelb and Bertschinger 1993 and references therein). On the other hand,
it would be interesting to study whether there might be a cooperative
analogue for velocity bias (Couchman and Carlberg 1992).

\bigskip
\noindent
{\large\bf Acknowledgements}

\bigskip

We thank Jim Fry for providing the data for Fig. 3 and L. N. da Costa
for providing the SSRS catalog.
This work was supported in part by DOE and by NASA (grant NAGW-2381)
at Fermilab. After this work was completed, we received a preprint
of Juszkiewicz, Bouchet, and Colombi which also gives some approximate
numerical results for $S_3(R)$ for standard (unbiased) CDM.

\newpage

\bigskip

\noindent
{\large\bf Figure Captions}
\bigskip

\noindent{\bf Fig. 1.} The two-point spatial correlation
function $\xi(r)/(b\sigma_8)^2$
in linear theory for standard CDM $(\Omega h = 0.5)$, low-density
CDM $(\Omega h =0.2)$, and CGF-modified standard CDM with $\kappa = 2.29$,
$R_s = 20 \Mpc$.

\bigskip

\noindent{\bf Fig. 2.} The volume-averaged normalized skewness
$S_3(R)$ in second-order
theory is shown as a function of top-hat smoothing radius $R$ for the
three models of Fig. 1.

\bigskip

\noindent{\bf Fig. 3.} The Fourier-space
3-point amplitude for equilateral triangles
$Q_\Delta (k)$ is shown as a function of wavenumber $k$. The data
points (from Baumgart and Fry 1991) are an average over subsamples from
the CfA and Perseus-Pisces surveys. The model points are for
standard perturbation theory (short dashed line, $Q_\Delta =4/7$),
and for the three CGF models discussed by BCFW: $\kappa, R_s =
0.84, 10 \Mpc$ (dot-long dash), $2.29, 20 \Mpc$ (solid), and
$4.48, 30 \Mpc$ (dot-short dash). For the
models, we have taken $b=1$, $c_2=0$.

\bigskip

\noindent{\bf Fig. 4.} The volume-average skewness
$S_3(R)$ for the same models as in Fig. 2 are shown in
comparison with data from the CfA and SSRS
surveys (from Gazta\~naga 1992). The models are shown with $b=1$,
$c_2 = 0$.

\bigskip

\noindent{\bf Fig. 5.} Contours of $\chi^2 = 5$, 8, and 14 (for 9 degrees of
freedom) in the $b-c_2$ parameter space for fits of the 3 models
to the data in Fig. 4. The darker regions correspond to lower
$\chi^2$. (a) CDM $\Omega h = 0.5$, (b) CDM $\Omega h = 0.2$, (c)
CGF $\kappa = 2.29$, $R_s = 20 \Mpc$.

\newpage

\bigskip

\noindent
{\large\bf References}
\bigskip

\def\pp{\par\parshape 2 0truecm 16.5truecm 1truecm 15.5truecm\noindent}
\def\paper#1;#2;#3;#4; {\pp#1, {#2}, {#3}, #4}
\def\book#1;#2;#3;#4; {\pp#1, {\sl #2} (#3: #4)}
\def\preprint#1;#2; {\pp#1, #2}

\paper Adams, F. C., Bond, J. R., Freese, K., Frieman, J. A., \&
Olinto, A. V. 1993;Phys.Rev.D;47;426;
\paper Bardeen, J. M., Bond, J. R., Kaiser, N., \& Szalay, A. S. 1986;
ApJ;304;15;
\paper Babul, A., \& White, S. D. M. 1991;MNRAS;253;31P;
\preprint Bardeen, J. 1984;%
in {\sl Inner Space/Outer Space}, eds. E. Kolb, M. Turner,
D. Lindley, K. Olive, and D. Seckel, (Chicago:Univ. of Chicago press, 1986);
\paper Baumgart, D. J., \& Fry, J. N. 1991;ApJ;375;25;
\paper Bernardeau, F. 1992;ApJ;392;1;
\preprint Bower, R., Coles, P., Frenk, C.S., \& White, S.D.M.
1993;ApJ;405;403;
\preprint Bouchet, F. R., Strauss, M., Davis, M., Fisher, K., Yahil, A.,
\& Huchra, J. 1993;preprint;
\preprint Bouchet, F. R., Davis, M., \& Strauss M. 1991;%
in {\sl The Distribution of Matter in the Universe}, eds. G. Mamon \&
D. Gerbal (Meudon: Observatoire de Paris);
\paper Bouchet, F. R. \& Hernquist, L. 1992;ApJ;400;25;
\paper Couchman, H. M. P. \& Carlberg, R. 1992;ApJ;389;453;
\paper Cen, R., Gnedin, N. Y., Kofman, L. A., \& Ostriker,
J. P. 1993;ApJ;399;L11;
\paper Cen, R., \& Ostriker, J. P. 1992; ApJ;399;L113;
\preprint Coles, P., Moscardini, L., Lucchin, F., Matarrese, S., \&
Messina, A. 1993;preprint;
\paper Da Costa, L. N., Pellegrini, P., Davis, M., Meiksin, A., Sargent,
W., \& Tonry, J. 1991;ApJS;75;935;
\paper Davis, M., Summers, F. J., \& Schlegel, D. 1992;Nature;359;393;
\paper Dekel, A., \& Rees, M. J. 1987;Nature;326;455;
\paper Efstathiou, G., Bond, J. R., \& White, S. D. M. 1992;MNRAS;258;1P;
\paper Efstathiou, G., Kaiser, N., Saunders, W., Lawrence, A.,
Rowan-Robinson, M., Ellis, R. S., \& Frenk, C. S. 1990;MNRAS;247;10P;
\paper Efstathiou, G., Sutherland, W., \& Maddox, S. J. 1990;Nature;348;705;
\preprint Feldman, H., Kaiser, N., \& Peacock, J. 1993; preprint UM AC 93-5;
\paper Fisher, K. B., Davis, M., Strauss, M. A., Yahil, A., \& Huchra, J. P.
1993;ApJ;402;42;
\paper Fry, J. N. 1984;ApJ;279;499;
\paper Fry, J. N. 1985;ApJ;289;10;
\paper Fry, J. N. 1986;ApJ;308;L71;
\preprint Fry, J.N. \& Gazta\~naga, E. 1993a;
ApJ in press (FERMILAB-Pub-92/367-A);
\preprint Fry, J.N. \& Gazta\~naga, E. 1993b;preprint FERMILAB-Pub-93/097-A;
\paper Fry, J. N. \& Seldner, M. 1982;ApJ;259;474;
\paper Gazta\~naga, E. 1992;ApJ;398;L17;
\preprint Gelb, J., and Bertschinger, E. 1993; preprint FERMILAB-Pub-92/74-A;
\paper Gelb, J., Gradwohl, B., \& Frieman, J. A. 1993;ApJ;403;L5;
\paper Goroff, M. H., Grinstein, B., Rey, S. J., \& Wise, M. B. %
1986;ApJ;311;6;
\paper Gramann, M., \& Einasto, J. 1991;MNRAS;254;453;
\paper Groth, E. J. \& Peebles, P. J. E. 1977;ApJ;217;385;
\paper Hamilton, A. J. S. 1988;ApJ;332;67;
\paper Hamilton, A. J. S., Kumar, P., Lu, E., \& Matthews, A. 1991;ApJ;374;L1;
\preprint Haynes, M., \& Giovanelli, R. 1988; in {\sl Large-Scale Motions
in the Universe}, ed. V. C. Rubin \& G. V. Coyne (Princeton: Princeton
University Press);
\paper Huchra, J., Davis, M., Latham, D., \& Tonry, J. 1983;ApJS;52;89;
\preprint Juszkiewicz, R., \& Bouchet, F. 1991;in {\sl The Distribution
of Matter in the Universe}, eds. G. Mamon \& D. Gerbal (Meudon: Observatoire
de Paris);
\paper Kaiser, N. 1984a;ApJ;284;L9;
\preprint Kaiser, N. 1984b; in {\sl Inner Space/Outer Space}, eds. E. Kolb,
M. Turner, D. Lindley, K. Olive, and D. Seckel,
(Chicago: University of Chicago press, 1986);
\paper Kaiser, N. 1987;MNRAS;227;1;
\paper Katz, N., Hernquist, L., \& Weinberg, D. H. 1992;ApJ;399;L109;
\preprint Katz, N., Quinn, P., \& Gelb, J. 1992;preprint;
\paper Lahav, O., Itoh, M., Inagaki, S., \& Suto, Y. 1993;ApJ;402;387;
\preprint Klypin, A., Holtman, J., Primack, J., \& Regos, E. 1992; preprint;
\preprint Liddle, A., \& Lyth, D. H. 1992;preprint;
\paper Liddle, A., Lyth, D. H., \& Sutherland, W. 1992;Phys.Lett.B;279;244;
\paper Loveday, J., Efstathiou, G., Peterson, B. A., \& Maddox, S. J. 1992;
ApJ;400;L43;
\paper Maddox, S. J., Efstathiou, G., Sutherland, W. J., \& Loveday,
J. 1990;MNRAS;242;43P;
\paper Meiksin, A., Szapudi, I., \& Szalay, A. S. 1992;ApJ;394;87;
\preprint Moscardini, L., Borgani, S., Coles, P., Lucchin, F.,
Matarrese, S., Messina, A., \& Plionis, M. 1993; preprint;
\paper Park, C., Gott, J. R., \& da Costa, L. N. 1992;ApJ;392;L51;
\paper Peacock, J. A., \& Nicholson, D. 1991;MNRAS;253;307;
\preprint Peebles, P. J. E. 1980; {\sl The Large Scale Structure of
the Universe}, (Princeton: Princeton University press);
\preprint Pogosyan, D., \& Starobinsky, A. 1992;preprint;
\paper Politzer, H. D., \& Wise, M. B. 1984;ApJ;285;L1;
\paper Rees, M. J. 1985;MNRAS;213;75P;
\paper Saunders, W., etal. 1991;Nature;349;32;
\paper Schaefer, R. K., Shafi, Q., \& Stecker, F. 1989;ApJ;347;575;
\paper Silk, J. 1985;ApJ;297;1;
\paper Szalay, A. S. 1988;ApJ;333;21;
\paper Szapudi, I., Szalay, A. S., \& Boschan, P. 1992;ApJ;390;350;
\paper Taylor, A. N., \& Rowan-Robinson, M. 1992;Nature;359;396;
\paper van Dalen, A., \& Schaefer, R. K. 1992;ApJ;398;33;
\paper Vittorio, N., Mattarese, S., \& Lucchin, F. 1988;ApJ;328;69;
\paper Vogeley, M. S., Park, C., Geller, M. J., \& Huchra, J. P. 1992;ApJ;395;
L5;
\paper White, S. D. M., Davis, M., Efstathiou, G., \& Frenk, C. S. 1987;
Nature;330;451;

\end{document}